\documentclass[journal]{IEEEtran}
\usepackage{amsmath}
\usepackage{bbm}
\usepackage{graphicx}
\usepackage{url}
\usepackage{algorithm,algorithmic}
\usepackage{amsfonts}
\usepackage{authblk}
\usepackage{color}
\usepackage{mathtools}  
\usepackage{amssymb}
\usepackage{tabulary}
\usepackage{booktabs}

\begin{document}
\title{Network Slicing for Vehicular Communication}
\author[1]{Hamza Khan}
\author[2]{Petri Luoto}
\author[1]{Sumudu Samarakoon}
\author[1]{Mehdi Bennis}
\author[1]{Matti Latva-aho}
\affil[1]{\{hamza.khan, sumudu.samarakoon, mehdi.bennis, matti.latva-aho\}@oulu.fi}
\affil[1]{Centre for Wireless Communications\\University of Oulu, FI-90014 Oulu, Finland}
\affil[2]{petri.luoto@mediatek.com}
\affil[2]{MediaTek Wireless Finland Oy\\Elektroniikkatie 16, Oulu }

\IEEEspecialpapernotice{(Invited Paper)}

\maketitle

\begin{abstract}
Ultra-reliable vehicle-to-everything (V2X) communication is essential for enabling the next generation of intelligent vehicles. V2X communication is a growing area of communication, that connects vehicles to  neighboring vehicles (V2V), infrastructure (V2I) and pedestrians (V2P). Network slicing is one of the promising technology for connectivity of the next generation devices, creating several logical networks on a common and programmable physical infrastructure. Network slicing offers an efficient way to satisfy the diverse use case requirements by exploiting the benefits of shared physical infrastructure. In this regard, we propose a network slicing based communication solution for vehicular networks. In this work, we model a highway scenario with vehicles having heterogeneous traffic demands. The autonomous driving slice (safety messages) and the infotainment slice (video stream) are the two logical slices created on a common infrastructure. We formulated a network clustering and slicing algorithm to partition the vehicles into different clusters and allocate slice leaders to each cluster. Slice leaders serve its clustered vehicles with high quality V2V links and forwards safety information with low latency. On the other hand, road side unit provides infotainment service using high quality V2I links. An extensive Long Term Evolution Advanced (LTE-A) system level simulator with enhancement of cellular V2X (C-V2X) standard is used to evaluate the performance of the proposed method, in which it is shown that the proposed network slicing technique achieves low latency and high-reliability communication.
\end{abstract}

\IEEEpeerreviewmaketitle

\section{Introduction}
Vehicle to vehicle and vehicle to infrastructure have become one of the key 5G vertical with use cases spanning from traffic safety to platooning. Today's vehicles are equipped with various intelligent systems i.e. automotive breaking system, cruise control and lane departure warning system. The vision of autonomous vehicles are based upon ultra reliable low latency communication (URLLC) enabling vehicles to communicate with their surrounding, and making well informed decisions. Vehicular communication includes three main categories: (i) vehicle-to-infrastructure (V2I) (ii) vehicle-to-vehicle (V2V) (iii) vehicle-to-pedestrian (V2P) all of which are collectively referred to as vehicle-to-everything (V2X) communication \cite{ultra}. 

Substantial number of accidents and fatalities occurring mostly at roundabouts and intersections signifies the need of communication between vehicles. Current research confirms human error as the cause of 90\% accidents \cite{national}. In the autonomous vehicle literature, V2X communication has been studied for more than a decade. Dedicated short range communication (DSRC), an IEEE 802.11p based technology, was considered as the key enabler for V2X communication in early 2000’s \cite{V2X_IEEE}. However, DSRC fails to guarantee the quality of service (QoS) requirements and suffers from unbounded latency \cite{unbounded} shifting the focus towards cellular V2X communication solution whose goal is to enable low-latency and ultra-reliable V2X communication \cite{V2X_LTE2}. 

The maximum end to end latency standardized by the European Telecommunication Standards Institute (ETSI) for a vehicular communication network is typically 100ms for a message size of about 1600 bytes along with the transmission reliability of 99.999\% \cite{speed}. These messages carry safety information such as cooperative cruise control, kinematics, vehicle location and collision warnings \cite{ref7}. 

\subsection{State-of-the-Art}
\begin{figure}[hbtp!]
	\centering
    \vspace{-5mm}
	\includegraphics[width=0.5\textwidth]{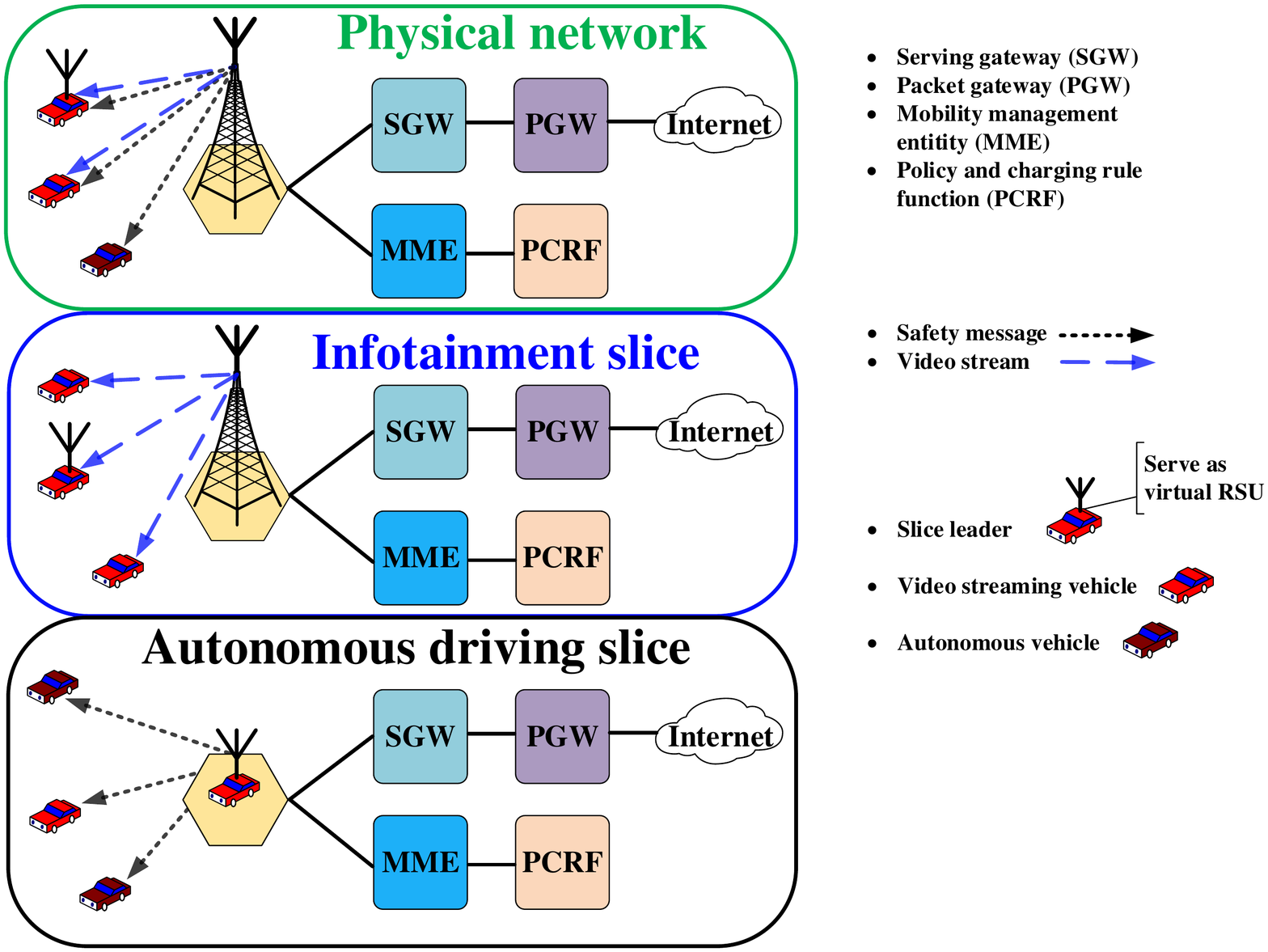}
	\caption{\label{Layout_Network}Illustration of network slicing for vehicular communication.}
\end{figure}

Radio resource allocation plays a central role to fulfill the QoS requirement of vehicular communication \cite{RRM}. Authors in \cite{mehdi} study the enablers of URLLC and respective tradeoffs. The end-to-end latency in vehicular networks is important from the perspective of reliability and it can be reduced by adopting non-orthognal multiple access (NOMA) technique for which authors in \cite{noma} proposed a centralized resource management solution. The applicability of NOMA in V2X services to achieve low latency and high reliability is studied in \cite{noma1}. The heterogeneous traffic demand of the next generation communication networks poses huge challenges in guaranteeing the quality-of-service (QoS) requirement of individual service. To investigate this, authors in \cite{resource_power} study the performance of joint resource allocation and power control for vehicle-to-vehicle (V2V) and vehicle-to-infrastructure (V2I) communication, where the QoS constraint for V2V guarantees high reliability, low latency and for V2I communication it guarantees the maximum sum rate. Authors in \cite{resource_1} studied the impact of resource allocation on cellular and vehicular users, where the resource allocation algorithm is designed as an optimization problem with strict latency and reliability constraints. The use cases of vehicular devices are increasing, while at the same time the limited cellular spectrum becomes a bottleneck in satisfying their QoS requirements. The unlicensed spectrum can be utilized to improve the performance of V2X networks and in this regard, authors in \cite{resource_2} study the coexistence problem of cellular V2X users and vehicular ad hoc network users over the unlicensed spectrum. On the other hand, to satisfy the stringent requirement of diverse V2X use cases network slicing is seen as a potential enabler \cite{network_slicing}. The aforementioned works \cite{noma}{ -}\cite{resource_2} assume a non-elastic/fixed network architecture, whereas authors in \cite{claudia} utilize network slicing to propose a flexible network architecture with service oriented functionalities. Radio resource allocation of different V2X network slices is studied in \cite{ran_slicing}, where the goal is to maximize the resource utilization and improve the system performance in terms of latency, achievable rate and outage. Moreover, authors in \cite{claudia_1} advocate a sliced network infrastructure as a prominent solution to support diverse V2X use cases and it is currently under discussion in 3GPP \cite{3gpp_NS}. The idea of network slicing is to logically isolate the different services from each other by creating multiple logical networks on a shared physical infrastructure \cite{slice1}. A sliced V2X communication network with independent QoS constraints is proposed in \cite{metis_slicing} to provide service to several use cases as illustrated in Fig. \ref{Layout_Network}. A network slice can span across all the network entities including the core network (CN) and radio access network (RAN), however in this work we have considered slicing in the RAN domain only. 

\subsection{Contribution}
The main contribution of this paper is to analyze the performance of a heterogeneous V2X communication network operating in the downlink direction. The heterogeneous services considered in this work include the safety messages and video streaming, which are part of the URLLC and the enhanced mobile broadband (eMBB) use case of 5G. To provide independent and isolated services, two sub-slices are created on the physical infrastructure named as autonomous driving slice and infotainment slice. Autonomous driving slice exchanges safety messages while infotainment slice provides video streaming services. In this work, we have considered a queuing system where the queues of video streaming packets are maintained at the road side unit (RSU) and the queues of safety message packets evolve at their respective slice leader (SL). The main goal of this paper is to propose a novel resource allocation solution for a sliced network considering the queue states information (QSI). In our proposed network slicing approach, the LTE-A RSU provides infotainment services only and clusters the vehicles (with weak QoS) and allocates them to SLs which will relay the data to clustered vehicles, where SLs act as autonomous driving slice access points. These access points act responsibly as virtual RSUs and exchange safety messages among the vehicles of autonomous driving slice.

This paper is structured as follows. Section II explains the system and link model. Network slicing for V2X communication is explained in Section III. Performance evaluation of the proposed technique is provided in Section IV. Concluding remarks are provided in Section V.

\emph{Notations}: We will use boldface lower case letter $\textbf{x}$ and boldface upper case letter $\textbf{X}$ to represent vectors and matrices respectively. The cardinality of set $\mathcal{X}$ is denoted as $X$ and $\textbf{x}^H$ denotes the Hermitian or the conjugate transpose of vector $\textbf{x}$.

\section{System Model}
This study considers single input multiple output (SIMO) transmission with orthogonal frequency division multiple access (OFDMA). The network components a set $\mathcal{B} = \{1,...,b,...,B\}$ of RSUs and a set $\mathcal{V}$ of vehicles. Vehicles in the network can be categorized as \emph{SLs} and \emph{clustered vehicles}. SLs or video streaming vehicles $\mathcal{S}$ have high quality V2I and V2V links and they serve as virtual RSU for their neighboring vehicles. Clustered vehicles $\mathcal{C}$ require safety messages only and they are served by the SLs via high quality V2V links. RSUs and SLs have their set of resources $\mathcal{M}^{\text{RSU}}$ and $\mathcal{M}^{\text{SL}}$ which consists of equal number of physical resource blocks (PRB). Each node in the network has $N_\text{r}$ receive antennas and $N_\text{t}$ transmit antennas. Vehicle partitioning in the network can be formally defined as:

\begin{subequations}
\label{sets}
\begin{align}
\label{set-partition} \mathcal{V} = {\mathcal{S} \cup \mathcal{C}} \hspace{10pt} \text{with} &\hspace{10pt} |\mathcal{V}| = |\mathcal{S}| +  |\mathcal{C}| \\
\mathcal{S}_b \cap  \mathcal{S}_{b'} = \emptyset \hspace{10pt} & \forall   b, b' \in \mathcal{B}, b \neq b' \\
\mathcal{C}_s \cap  \mathcal{C}_{s'} = \emptyset \hspace{10pt} & \forall   s, s' \in \mathcal{S}, s \neq s' \\
\end{align}
\end{subequations}

The communication channel from RSU $b$ to SL $s$ over subcarrier $m \in \mathcal{M}^{\text{RSU}}$ is denoted as $\textbf{h}_{b,s}^{m}(t)$. Similarly, the channel vector from autonomous slice access point $s$ to vehicle $c$ is denoted as $\textbf{h}_{s,c}^{m}(t)$, where $ m \in \mathcal{M}^{\text{SL}}$. The received signal by video streaming vehicle at time $t$ is given by (\ref{receivedsignalvideo}) and the received signal by autonomous vehicle is given by (\ref{receivedsignalsafety}):

	\begin{equation}
	\textbf{y}_{b,s}^{m}(t) = \textbf{h}_{b,s}^{m}(t)\textbf{x}_{b,s}^{m}(t)+\sum_{b'\in \mathcal{B}\setminus b}\textbf{h}_{b',s}^{m}(t)\textbf{x}_{b',s'}^{m}(t)+\textbf{z}_{b,s}^{m}(t)
	\label{receivedsignalvideo},
	\end{equation}
	
	\begin{equation}
	\textbf{y}_{s,c}^{m}(t) = \textbf{h}_{s,c}^{m}(t)\textbf{x}_{s,c}^{m}(t)+\sum_{s'\in \mathcal{S}\setminus s}\textbf{h}_{s',c}^{m}(t)\textbf{x}_{s',c'}^{m}(t)+\textbf{z}_{s,c}^{m}(t)
	\label{receivedsignalsafety},
	\end{equation}
where $\textbf{x}_{b,s}^{m}(t)$ is the transmitted signal from the RSU $b$ to SL $s$,  $\textbf{h}_{b',s}^{m}(t)$ is the channel vector from interfering RSU to the \emph{s}-th vehicle, and $\textbf{z}_{b,s}^{m}(t)$ is the additive noise. Similarly, $\textbf{x}_{s,c}^{m}(t)$ is the transmitted signal from access point $s$ to autonomous vehicle $c$ and the interfering channel vector is $\textbf{h}_{s',c}^{m}(t)$. Interference in the network occurs when resources are reused by multiple RSUs or SLs.

The network model assumes that each vehicle has queues of video frames maintained at their serving RSU and with fixed arrival rate. The evolution of queue of $s$-\text{th} vehicle at the RSU $b$ is:
\begin{align}
q_{b,s}(t+1) = [q_{b,s}(t) - r_{b,s}(t)]^{+} + r_{s}^{\text{Arrival}}  & \quad s \in \mathcal{S},
\end{align}
where $[\cdot]^+$ indicates the max $\{0,\cdot\}$, $r_{b,s}(t)$ is the service rate of RSU $b$ and $r_{s}^{\text{Arrival}}$ is the packet arrival rate of $s$-\text{th} vehicle under the assumption of deterministic periodic arrivals. The queuing system considered in this work can be categorized as D/G/1. The service rate of the queuing system depends upon the capacity of the channel and the queue length i.e., $r_{b,s} = \min(q_{b,s},r_{b,s})$. On the other hand, if we consider a system without queues then the service rate depends only on the channel capacity.

The eMBB slice consists of RSU's and the vehicle's requesting video streaming services. Due to the nature of video streaming service, continuous transmission of video packets are required to minimize the fluctuations of video quality during video playback. In this regard, we have used a continuous packet arrival rate of 1000 bits/ms for the eMBB slice, where the packets intended for video streaming vehicle are queued at the RSU.

On the other hand, the URLLC traffic is generated by the safety information sharing among vehicular users. The safety messages consist of periodic transmission containing information of vehicle location, kinematics and collision warnings, etc, to ensure passenger and vehicle safety. These messages are of fixed packet size 1280 bits and are transmitted after every 10ms for each autonomous vehicle. The reliability of safety message transmission is of utmost importance as the packet loss or re-transmission can lead to accidents due to the non-availability of the latest information. The reliability of 99.999\% has to be ensured for URLLC transmissions \cite{99cite}. The autonomous driving slice vehicles which receive safety messages from SLs have queues evolving at their serving SL $s \in \mathcal{S}$ given as:

\begin{align}
q_{s,c}(t+1) = [q_{s,c}(t) - r_{s,c}(t)]^{+} + r_{c}^{\text{Arrival}} & \quad c \in \mathcal{C},
\end{align}
where $r_{s,c}(t)$ is the service rate of SL and $r_{c}^{\text{Arrival}}$ denotes the arrival of safety messages for clustered vehicle $c$. The arrival of safety messages is also deterministic with a periodic arrival interval of 10ms.

\begin{figure}[hbtp!]
	\centering
	\includegraphics[width=0.48\textwidth]{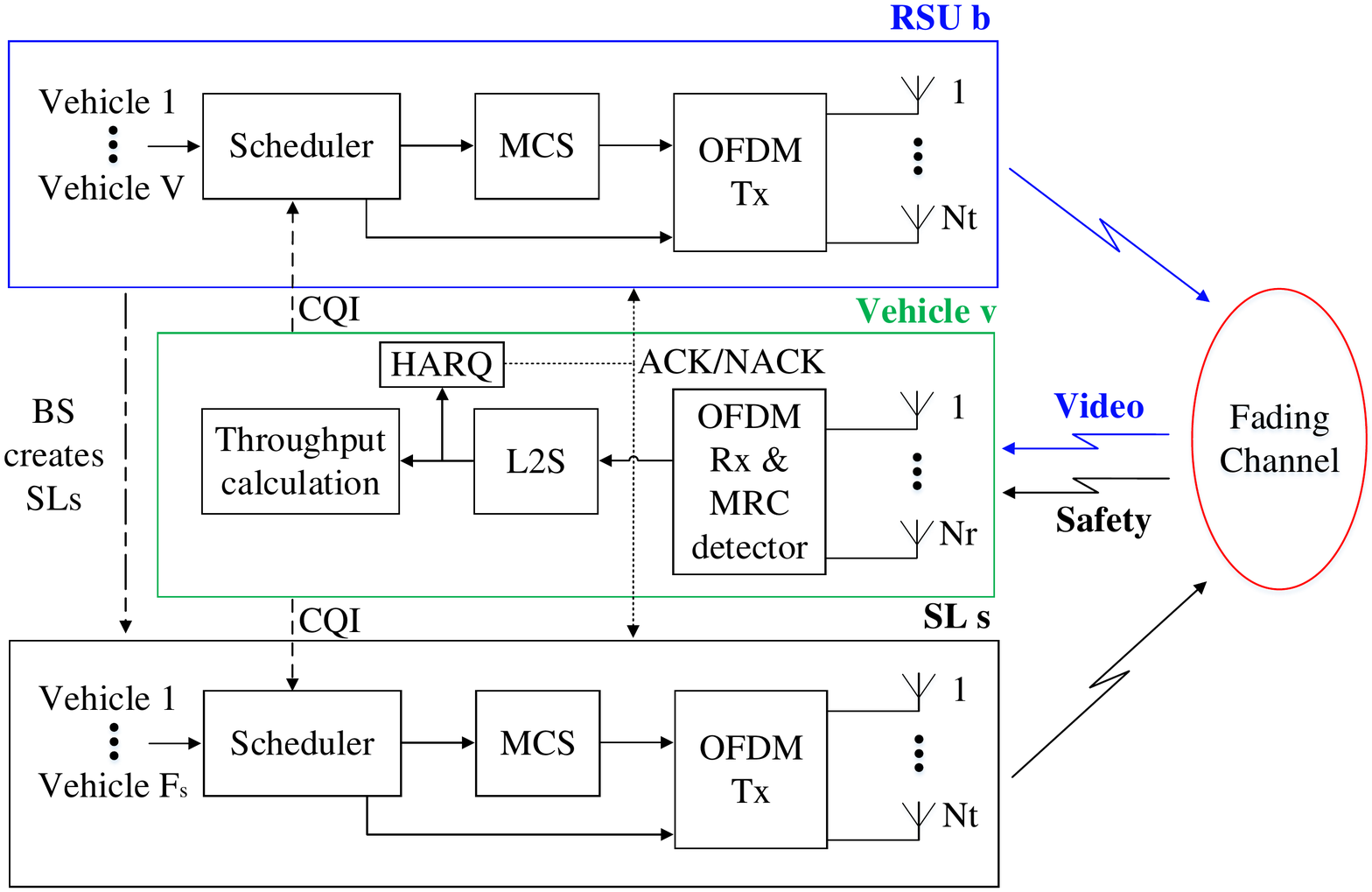}
	\caption{\label{Link}Wireless link modelling of the sliced vehicular network.}
\end{figure}

A geometry-based stochastic channel model (GSCM) \cite{WINII} is used to model a wireless link between SL and the RSU. Independent and identity distributed (i.i.d) channel is anticipated for the vehicles of autonomous driving slice which follow the path loss model of \cite{V2V_pathloss} reflecting the macro to relay path loss model. Wireless link modelling is illustrated in Fig. \ref{Link}. 

Radio links are modelled with link-to-system (L2S) interfaces for which channel quality information (CQI) is considered as the most substantial parameter. Each vehicle transmits the CQI information which is used to estimate the modulation and coding schemes (MCS). Afterwards, proportional fair (PF) scheduling algorithm is employed by the RSU and slice access points to utilize the network resources efficiently. PF scheduler maximizes the throughput and pledges to provide minimum service threshold to all the vehicles. In the process, autonomous driving slice access points schedule safety messages for all the vehicles, while RSU schedule only those vehicles requiring infotainment services. The effects of the inter-symbol interference (ISI) are ignored as the cyclic prefix is assumed to be larger than the delay spread.

\begin{table}[hbtp!]\caption{Table of Notation}
\begin{center}
\begin{tabular}{>{}l >{}c p{6.5cm}}
\toprule
\multicolumn{3}{c}{\textbf{Sets}} \\
\hline $\mathcal{B}$ & $\triangleq$ & Set of RSUs\\
$\mathcal{V}$ & $\triangleq$ & Set of Vehicles\\
$\mathcal{S}$ & $\triangleq$ & Set of SLs\\
$\mathcal{C}$ & $\triangleq$ & Set of Clustered vehicles\\
$\mathcal{\mathcal{M}^{\text{RSU}}}$ & $\triangleq$ & Set of RSU resources\\
$\mathcal{\mathcal{M}^{\text{SL}}}$ & $\triangleq$ & Set of SL resources\\  \hline 
\multicolumn{3}{c}{\textbf{System Model Variables}}\\
\hline $\textbf{h}_{b,s}^{m}$ & $\triangleq$ & Channel from RSU $b$ to SL $s$ over subcarrier $m \in \mathcal{M}^{\text{RSU}}$\\
$\textbf{h}_{s,c}^{m}$ & $\triangleq$ & Channel from SL $S$ to clustered vehicle $c$ over subcarrier $m \in \mathcal{M}^{\text{SL}}$\\
$\textbf{x}_{b,s}^{m}$ & $\triangleq$ & Transmitted signal from the RSU $b$ to SL $s$\\
$\textbf{x}_{s,c}^{m}$ & $\triangleq$ & Transmitted signal from SL $s$ to clustered vehicle $c$\\
$\textbf{y}_{b,s}^{m}$ & $\triangleq$ & Received signal from the RSU $b$ to SL $s$\\
$\textbf{y}_{s,c}^{m}$ & $\triangleq$ & Received signal from SL $s$ to clustered vehicle $c$\\
$q_{b,s}$ & $\triangleq$ & Evolution of vehicle $s$ queue at the RSU $b$\\
$r_{b,s}$ & $\triangleq$ & Service rate of RSU $b$\\
$r_{s}^{\text{Arrival}}$ & $\triangleq$ & Packet arrival rate of $s$-\text{th} vehicle\\
$q_{s,c}$ & $\triangleq$ & Evolution of vehicle $c$ queue at the SL $s$\\
$r_{s,c}$ & $\triangleq$ & Service rate of SL $s$\\
$r_{c}^{\text{Arrival}}$ & $\triangleq$ & Packet arrival rate of $c$-\text{th} vehicle\\
\hline \multicolumn{3}{c}{\textbf{Clustering Variables}}\\ \hline
$\textbf{A}$ & $\triangleq$ & Euclidean distance based similarity matrix\\
$\textbf{W}$ & $\triangleq$ & Weighted adjacency matrix of $\textbf{A}$\\
$\textbf{L}$ & $\triangleq$ & Laplacian matrix\\
$\textbf{D}$ & $\triangleq$ & Degree martix \\
$\textbf{U}$ & $\triangleq$ & Matrix of eigenvectors of $\textbf{L}$ \\
$d_{vv'}$ & $\triangleq$ & Gaussian similarity function for vehicles $v,v' \in \mathcal{V}$\\
\bottomrule
\end{tabular}
\end{center}
\label{tab:TableOfNotationForMyResearch}
\end{table}

To exploit spatial diversity, maximal ratio combining (MRC) detector is deployed at the receiving node. The receiving nodes perform SINR calculations at each physical resource block (PRB). It is assumed that symbol are perfectly synchronized in both time and frequency domains. The exact modelling of radio links cause computational overhead which is abridged by utilizing mutual information based effective SINR (MIESM) mapping. SINR is mapped to the corresponding information curves with the help of MIESM. 

The evaluation of erroneous transmission then takes place by mapping mutual information to the frame error curves of conforming MCS. Successful transmission results in an acknowledgement (ACK), whereas failed transmissions effects in a negative acknowledgements (NACK). In case of failed transmissions, a hybrid automatic repeat request (HARQ) is generated to retransmit the unsuccessful frames.

\section{Vehicle Clustering Via Network Slicing}
Network slicing offers an efficient way to satisfy the diverse use case requirements of the next generation vehicular devices. It is defined as the concept of creating multiple logical networks on a shared physical infrastructure. Network slicing allows the operator to flexibly provide dedicated services over a common infrastructure. Moreover, network slicing brings elasticity, flexibility and scalability, which will be helpful in addressing the issues of efficiency for the future generation of mobile services and it is illustrated in Fig. \ref{Layout_Network}. The key enabler of this technology is network functions virtualization (NFV) and software defined networking (SDN) \cite{sdn}, which can reconfigure networks with software and manage network resources. NFV allows the placement of network functions in convenient locations i.e., user plane/control plane functionalities can be moved close to the end user to reduce the service latency. The proposed network slicing approach involves the partitioning of the radio access network along with the configuration of vehicular devices functionality to support the different use cases. In this regard, the video streaming vehicles in our work reconfigure their functionality to serve the clustered vehicles in its neighborhood. The network functions are reconfigured via software, which involves the assignment of new SLs and their serving vehicles. In this way, SDN and NFV perform the reconfiguration of serving nodes and the placement of RSU functionality in the reconfigured nodes. The physical network in Fig. \ref{Layout_Network} represents the legacy LTE-A network where the RSU is responsible for every service. When this physical network is divided into slices the service function chain of the infotainment slice remains the same i.e., the RSU, while the service function chain for the autonomous driving slice changes to the SL. The SLs are reconfigured with the help of NFV, as these network functions are not tied to the underlying hardware they can be deployed as virtual RSUs. The virtual network functions can be relocated, scaled and dynamically instantiated based on the network demand and the  environment dynamics. 

Network slicing enables the reconfiguration of service functions via NFV, with which the same infrastructure can be used to provide different services. The key step in network slicing is to determine clusters of vehicles. RSU measures the CQI value of all the vehicles and computes the similarity between the video streaming vehicles $\mathcal{S}$ and autonomous driving vehicles $\mathcal{C}$ in the network based on the geographical information. RSU constructs the distance based similarity matrix \textbf{A} in a way that vehicles which have large distance have less similarity and vice versa. Since the decoupled scheduling requires the partitioning of set $\mathcal{V}$ as per (\ref{sets}), we decouple the SL and clustered vehicles and first utilize spectral clustering algorithm to determine $\mathcal{C}$ \cite{clustering_tutorial}.

\begin{algorithm}
 \caption{Partitioning clustered vehicles}
 \label{clustering}
 \begin{algorithmic}[1]
 \renewcommand{\algorithmicrequire}{\textbf{Input:}}
 \renewcommand{\algorithmicensure}{\textbf{Output:}}
 \REQUIRE Similarity matrix \textbf{A}
 \ENSURE  a set of k clusters i.e. $\cup \mathcal{C}_k$
 \\ \textit{Initialization}:
  \STATE Let \textbf{W} be the weighted adjacency matrix of \textbf{A}.
  \STATE Compute the unnormalized Laplacian i.e \textbf{L} = \textbf{D} - \textbf{W}, where \textbf{D} is a degree matrix of \textbf{A}.
  \STATE The number of clusters k corresponds to the index of maximum eigenvalue of \textbf{L}.
  \STATE Let \textbf{U} $\in \mathbb{R}^{n\text{x}k} $ represent the k eigenvectors of \textbf{L} i.e. $u_1, ... , u_k$ corresponds to the columns of \textbf{U}.
  \FOR {$i = 1,...,n$}
  \STATE let $x_i \in \mathbb{R}^k$ be the vector representing the $i$-th row of \textbf{U}.
  \STATE Cluster the points $(x_i)_{i=1,...,n}$ in $\mathbb{R}^k$ with the k-means algorithm into clusters $\mathcal{C}_1,..., \mathcal{C}_k$.
  \ENDFOR
 \RETURN $\mathcal{C} = \{\mathcal{C}_1,...,\mathcal{C}_k \}$ 
 \end{algorithmic} 
 \end{algorithm}

Euclidean distance based similarity matrix $\textbf{A}$ is formulated at each RSU, where $d_{vv'}$ corresponds to the $(v,v')$-th entry. Similarity \textbf{A} between two vehicles $v,v' \in \mathcal{V}$ is measured by the Gaussian similarity function given as \cite{gaussian}: 

\begin{equation}
d_{vv'} = \exp \bigg( \frac{-||{d}_{v} - {d}_{v'}||}{2\sigma^{2}} \bigg)
\label{receivedsignal},
\end{equation}
where $d_v$ corresponds to the location of vehicle $v$. Impact of neighborhood size is controlled by $\sigma$. Small value of $\sigma$ results into less vehicles per cluster and vice versa. Apart from $\sigma$, an important clustering parameter is the input $k$ which determines the number of clusters. If $\sigma$ is fixed, an appropriate choice of clusters $k$ is determined by eigenvalue method. Spectral clustering exploits the geometry of nodes in a graph. The graph Laplacian matrix is formulated as \textbf{L} = \textbf{D} $-$ \textbf{A}, where \textbf{D} is diagonal matrix with $v$-th diagonal element given as $\sum_{v'=1}^{V} d_{vv'}$. For cluster $\mathcal{C}$, the combination of the smallest $\mathcal{C}$ eigenvectors of \textbf{L} can be used to determine the input parameter $k$ for $k$-means clustering. 

\begin{equation}\label{similarity}
k = \arg\max_{i}(\chi_{i+1}-\chi_i), \hspace{0.2cm} i = 1,...,n-1,
\end{equation}
where $\chi_i$ is the $i$-th smallest eigenvalue. When the nodes on the graph are uniformly distributed as $k$ clusters, the first $k$ eigenvalues are small and the ($k+1$)th eigenvalue becomes relatively large. Since the spectral clustering algorithm requires the knowledge of similarity matrix \textbf{A}, it is categorized as a centralized clustering mechanism. The output of clustering algorithm is the set of clustered vehicles $\mathcal{C} = \{ \cup \hspace{2pt} \mathcal{C}_k \}$.
\begin{algorithm}
 \caption{Slicing Algorithm}
 \label{slicing}
 \begin{algorithmic}[1]
 \renewcommand{\algorithmicrequire}{\textbf{Input:}}
 \renewcommand{\algorithmicensure}{\textbf{Output:}}
 \REQUIRE {$\mathcal{C}$}
 \ENSURE  $\mathcal{S}$
 \\ \textit{Initialization}:
  \FOR {each cluster $i = 1,...,k$}
  \STATE Let $y_i$ represent the center of cluster $i$
  \STATE Find the distance between $y_i$ and vehicles $v \in \mathcal{V} \backslash \mathcal{C}$
  \STATE Find $s_i = \text{arg min}_{\forall v \in \mathcal{V} \backslash \mathcal{C}}\{y_i - d_v\}$, where $d_v$ is the location of vehicle $v$
  \ENDFOR
 \RETURN $\mathcal{S} = \{s_1,...,s_k\}$ 
 \end{algorithmic} 
 \end{algorithm}

To serve the set of clustered vehicles we need to determine the SLs $\mathcal{S}$. As described earlier, $\mathcal{S}$ is the set of vehicle which have high quality V2I link and V2V link, which makes it suitable for them to serve as an access point. RSU utilizes the geographical information to determine the distance of all vehicles $v\in \mathcal{V} \backslash \mathcal{C} $ from cluster center. Since the vehicular devices are mobile, it is highly likely that the nearest vehicle from cluster center will have high quality V2V links with all the vehicles of the respective cluster. A vehicle which is the closest to the cluster center then becomes the SL as per the slicing algorithm. After determining the clustered vehicles $\mathcal{C}$ and SLs $\mathcal{S}$, RSU schedules the SL, and the SL respectively schedules the clustered vehicles.  In the upcoming section a performance analysis of the vehicular network with different network densities is performed.

\begin{figure}[hbtp]
	\centering
	\includegraphics[width=0.48\textwidth]{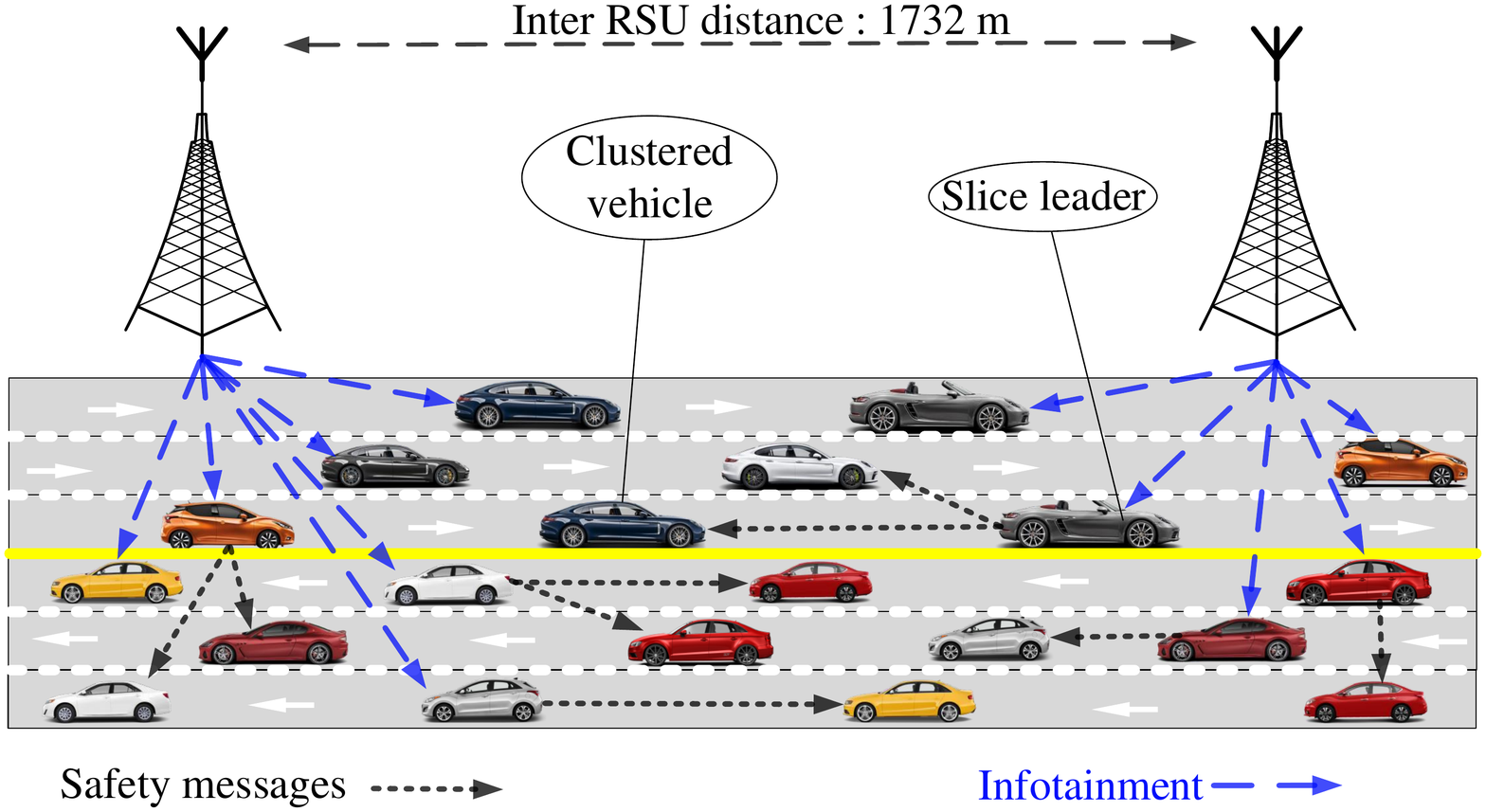}
	\caption{\label{Layout}Layout of the sliced vehicular network.}
\end{figure}

\section{Performance Analysis}
System level simulators provide accurate analysis of the anticipated performance. Simulation results provide a better insight of various issues i.e. throughput analysis, outage analysis, performance of the scheduling algorithm and interference management. This work analyzes the performance of network slicing based cellular V2X network with two sub-slices. Simulation scenario considers a six lane highway layout as shown in Fig. \ref{Layout} and earlier analyzed in \cite{hamza}. Vehicles in three lanes are moving in the left direction and vehicles in the remaining three lanes are moving towards the right direction. Vehicles on the highway are moving with a speed of 140 km/h. The RSU network is situated along the highway at a distance of 35 m from the first lane and the inter-RSU distance is 1732 m. 
	
Autonomous vehicles are moving in each lane of the highway over a 10 km long stretch. Various vehicular densities are modelled by changing the inter-vehicular distance. Sparse network is simulated by keeping the inter-vehicular distance large and dense network is simulated by keeping the inter-vehicular distance small. The network slicing procedure mentioned in algorithms \ref{clustering}, \ref{slicing} is repeated by the RSU after every 100ms to update the V2I and V2V links. The clustering algorithm ensures that access points only serve those autonomous vehicles with which it can guarantee reliable communication.

SLs forward the safety information to the vehicles that are clustered within the same zone. The RSU forwards infotainment packets to video streaming vehicles. The RSU communicates with the vehicles in the downlink direction with a transmit power of 46 dBm and the transmission packet size is 1000 bits with a scheduling interval of 1ms. Moreover, SL communicates with the clustered vehicles with transmission power of 20 dBm for a message size of 1280 bits and the periodicity is 10ms. 

\begin{table}[hbtp]
	\centering
	\caption{Simulator parameters.}%
	\label{params}
	\begin{tabular}{p{3cm} p{4.2cm} }
		\toprule
		\textbf{Parameter} & \textbf{Value} \\ \hline \hline
		Duplex mode & FDD \\ 
		System Tx bandwidth & 10 MHz \\ 
		Antenna configuration & 1 Tx $\times$ 2 Rx\\
		Receiver type & MRC \\ 
		Vehicle speed & 140 km/h \\ 
		Scheduler & Proportional fair \\ 	
		L2S interface metric & MIESM \\ 		
		Synchronization & Time and frequency synchronized \\ 
		HARQ & Chase combining \\ 
		Inter RSU distance & 1732 m\\ 
		& Scenario 1:  \hspace{8pt}1-100 m, \\
		Inter-vehicular Distance & Scenario 2: 100-200 m, \\
		& Scenario 3: 200-300 m \\ 	
		\hline \multicolumn{2}{c}{ \textbf{LTE-Uu (V2I)} } \\ \hline
        Carrier Frequency      & 2 GHz \\ 
        Transmission Power  & 46 dBm    \\ 
        Throughput   & 1000 kbits/s     \\ 
        \hline \multicolumn{2}{c}{ \textbf{LTE-PC5 (V2V)} } \\ \hline 
        Carrier Frequency      & 5.9 GHz \cite{3GPP_frequency}\\ 
        Transmission Power  & 20 dBm    \\ 
        Throughput   & 128 kbits/s     \\ 
        \bottomrule
	\end{tabular}
\end{table}

In this work, the target communication technology is C-V2X. The C-V2X standard includes two radio interfaces named as Uu and PC5 \cite{Cv2x}. The legacy LTE-A transmission uses the LTE-Uu interface for the provisioning of all services. However, when we slice the network we utilize the LTE-Uu interface for the infotainment slice and the LTE-PC5/sidelink interface for the autonomous driving slice. Autonomous driving slice uses frequency of 5.9 GHz dedicated by 3GPP for the PC5 interface, while the infotainment slice uses frequency of 2 GHz. The sidelink interface can operate in network scheduled mode (RSU schedules the resources) or the autonomous mode (a resource pool is configured and vehicle autonomously schedules the resources without the involvement of RSU). In this work, we have used the autonomous mode for the scheduling of safety messages. The autonomous mode can operate without cellular coverage and it is considered as a baseline V2V transmission mode due to the fact that safety application cannot be restricted only to the areas with cellular coverage.

The rest of simulation parameters are summarized in Table \ref{params}. We analyze the performance of the proposed network slicing algorithm in terms of throughput, latency, queuing dynamics and the packet failure ratio. The performance of the proposed method is compared with the state-of-the-art fixed service chain architecture utilized by LTE-A and the performance gain highlights the importance of using flexible networks in the future. The first baseline degrades the performance of the cell edge vehicle as they have poor SINR so a second baseline is introduced for comparison which utilizes relaying mechanism for the low SINR cell-edge vehicles.

\subsection{Network Slicing}\label{sec:NS} 
In this section we analyze the performance of the proposed network slicing algorithm. A brief description of the compared baselines are provided in Table \ref{description}. Baseline 1 in this work is representing the legacy LTE-A network which is inelastic and the RSU schedules users for variety of services. In the currently deployed network architecture serving the nodes at the cell edge is a challenging task and to overcome this problem author in \cite{Petri2} proposed a relaying technique for the nodes at the cell edge. Baseline 2 in this work represents the relaying approach where low SINR cell edge vehicles are relayed. The proposed network slicing algorithm with variable neighborhood size is compared with both the baselines under different network densities.  

\begin{table}[hbtp]
	\centering
	\caption{Description of baselines.}%
	\label{description}
	{
		\begin{tabular}{{|p{1cm}|p{6.2cm}|}}
			\hline
			\textbf{Baselines} & \textbf{Description} \\
			\hline
			1 & RSU provides coverage in a fixed radius. Vehicles at cell edge suffer from performance degradation. RSU is providing multiple services without isolation. Service function chains are predefined and cannot be re-configured. The resources of RSU are shared among all the users in a proportional manner. \\ \hline
			2 & Coverage is provided in a similar way as baseline 1 with the addition of an offloading mechanism for vehicles with low SINR. The offloaded vehicles are served by others which have high quality V2I and V2V link. RSU assigns the offloaded vehicle to access point located close to it. The available resources at the RSU are shared by the directly connected vehicles.\cite{Petri2} \\ \hline

	\end{tabular}}
\end{table}

The cumulative distribution function (CDF) of the throughput and queue state is analyzed with the complementary cumulative distribution function (CCDF) of queuing latency. The colors in the figure represent different vehicular densities i.e., red color represents dense network with vehicular spacing of 1-100 m, green represent inter-vehicular distance of 100-200m and blue color is used for sparse network with inter-vehicular distance 200-300 m. The network slicing algorithm proposed in this work creates two sub-slices isolated from each other. The infotainment sub-slice includes the RSU and the video streaming vehicles, while the autonomous driving sub-slice includes all the vehicles that are served by their SL in the downlink direction. In this way the resources of RSU and SL are utilized to provide individual services hereby isolating the two sub-slices. 

The first algorithm in the network slicing method cluster the vehicles into $k$ clusters based on their similarity of geographical information. Each cluster of vehicle is served by a SL and is chosen as per the slicing algorithm described in Section (III), which chooses the vehicle closest to the cluster center as the SL. Gaussian similarity function defined in \eqref{receivedsignal} determines the similarity between the SL and the clustered vehicles and neighborhood size parameters play an important role. The choice of neighborhood size parameter comes with a trade off between creating large number of SL which will increase the interference and putting more vehicle in one cluster which cannot guarantee that all the vehicles will be scheduled at all time slots, since the proportional fair scheduler distributes the resources in a proportional manner. 

The proposed algorithm creates several SLs on the basis of neighbourhood size parameter. In this work we have used neighborhood size of $5\hspace{2pt} \& \hspace{2pt}50$ m, where smaller neighborhood size reflects less similarity and vice-versa. When $\sigma = 5$ m the number of SLs in the network are as high as $S = 115$ for the dense network and the number of SL decreases as the network becomes sparser or the neighborhood size is increased. On the other hand, when $\sigma = 50$ m the similarity between vehicle increases and we get $S = 10$ SLs which are 10 times less compared to smaller $\sigma$ for the same neighborhood size. 

\begin{figure}[hbtp]
	\centering
	\includegraphics[width=0.5\textwidth]{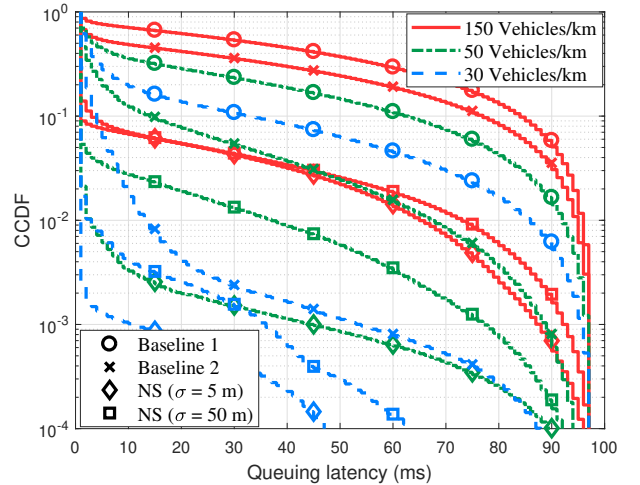}
	\caption{\label{Latency_autonomous} CCDF of the queuing latency of autonomous driving sub-slice.}
\end{figure}

\begin{figure}[hbtp]
	\centering
	\includegraphics[width=0.5\textwidth]{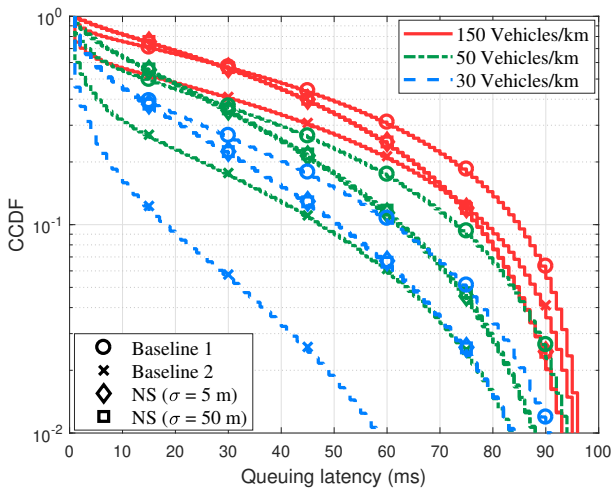}
	\caption{\label{Latency_infotainment} CCDF of the queuing latency of infotainment sub-slice.}
\end{figure}

Now we analyze the performance of the proposed method with the above mentioned baselines. Fig. \ref{Latency_autonomous} plots the CCDF of the queuing latency of autonomous driving slice and Fig. \ref{Latency_infotainment} shows the CCDF of latency of infotainment slice. Queuing latency reflects the time a packet spends in the queue before scheduling. The worst case queuing latency is observed in the dense network for the baseline 1 as all the vehicles and all the services are served by the RSU. From the figures we can see that the queuing latency decreases with the increase of inter-vehicular distance i.e. the sparse network experiences the lowest latency. The baseline 2 achieves slightly better performance than baseline 1 for the vehicle of autonomous driving slice for all network density as cell-edge vehicles are offloaded to neighboring vehicles. Next we analyze the performance of the proposed method for the autonomous slice as reliable transmission of safety messages is of huge importance. From Fig. \ref{Latency_autonomous} we can see that baselines 1 and 2 perform the worst in all network densities and the proposed algorithm achieves significant performance gains as compared to both the baselines. The proposed method utilizes high quality V2V and V2I links for the reliable transmission of safety messages rather than utilizing poor quality V2I link. Moreover, the impact of neighborhood size is visible from the Fig. \ref{Latency_autonomous} where smaller neighborhood size results into less queuing delay and large value of $\sigma$ results in higher queuing delay. When $\sigma$ is large we have less number of cluster leaders and ensuring high quality V2V link in this case becomes challenging as far away vehicles can lie within the same cluster due to less value of similarity. On the other hand small value of $\sigma$ creates large number of access points which ensure the V2V link quality. 

Similarly, Fig. \ref{Latency_infotainment} illustrates the performance of the baselines and the proposed network slicing algorithm for the infotainment sub-slice. The performance of the network slicing algorithm is a compromise between baseline 1 and 2, where baseline 1 performs the worst for all the network densities. Baseline 2 performs better than the rest because of relaying the low SINR vehicles which offloads these weak V2I link vehicles to high quality V2V links. Neighborhood size of the proposed method doesn't effect the performance of infotainment slice since the SLs are only responsible for the exchange of safety messages. 

\begin{figure}[hbtp]
	\centering
	\includegraphics[width=0.5\textwidth]{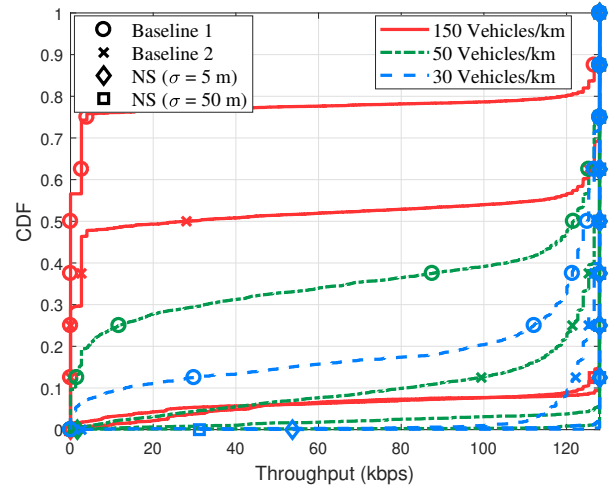}
	\caption{\label{Throughput_autonomous} CDF of the throughput of autonomous driving sub-slice.}
\end{figure}

\begin{figure}[hbtp]
	\centering
	\includegraphics[width=0.5\textwidth]{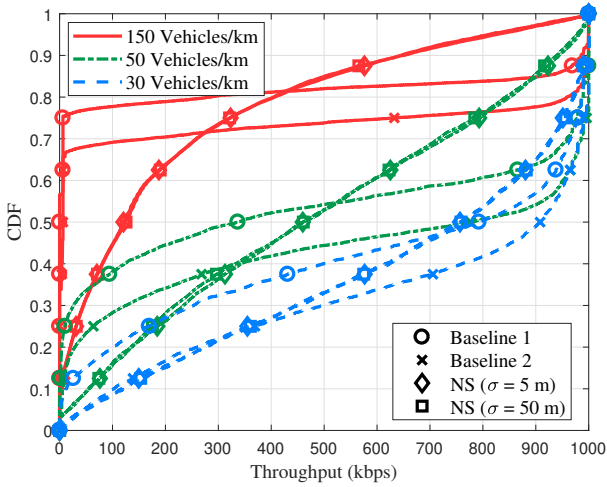}
	\caption{\label{Throughput_infotainment} CDF of the throughput of infotainment sub-slice.}
\end{figure}

Moreover we also analyze the performance of the proposed method by comparing the CDF of throughput of autonomous sub-slice shown in Fig. \ref{Throughput_autonomous} and infotainment sub-slice plotted in Fig. \ref{Throughput_infotainment} with the baselines. The throughput of the autonomous slice performs the best when network slicing is introduced. The proposed method achieves the maximum throughput when $\sigma = 5$ m i.e. 99.47 \% of the vehicles achieve the target throughput of 128 kbps. The dense network shows 70 \% increase in the number of vehicles in terms of the achievable throughput for the proposed method when compared to the baseline 1. The maximum number of vehicles achieve the target value of throughput in case of the sparse network with inter-vehicular distance of $200-300$ m. Furthermore, the infotainment sub-slice throughput shown in Fig. \ref{Throughput_infotainment} shows huge improvement for the baseline 2 compared to the proposed method and the baseline 1. The proposed method increase the number of users which are experiencing non-zero throughput compared to baseline 1. On the other hand, the baseline 2 tries to push the maximum achievable throughput by all the vehicles by offloading vehicles which are below a certain SINR threshold. Baseline 2 maintains the link quality by ensuring the SINR threshold and by only utilizing the high quality V2I and V2V links baseline 1 provides the best performance for all the network densities.

\begin{figure}[hbtp]
	\centering
	\includegraphics[width=0.5\textwidth]{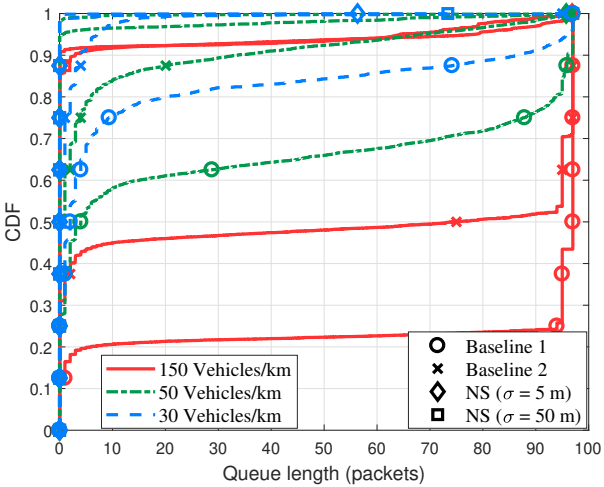}
	\caption{\label{Queue_autonomous} CDF of the queue length of autonomous driving sub-slice.}
\end{figure}

\begin{figure}[hbtp]
	\centering
	\includegraphics[width=0.5\textwidth]{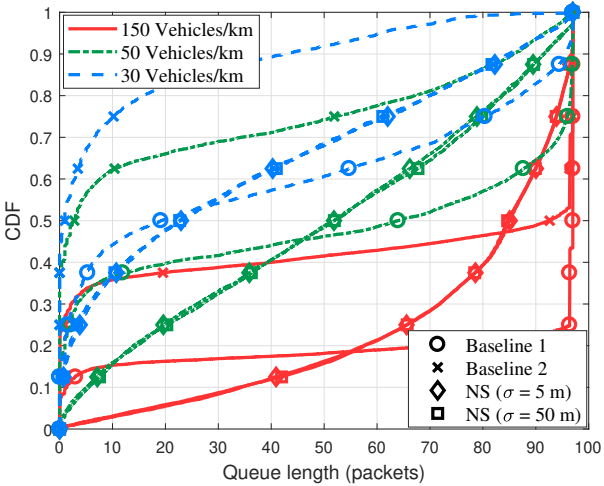}
	\caption{\label{Queue_infotainment} CDF of the queue length of infotainment sub-slice.}
\end{figure}

Finally, we analyze the CDF of the packet queue length at the RSU and the SL. The CDF of the queue length of the autonomous driving sub-slice is shown in Fig. \ref{Queue_autonomous} where the queues of safety messages are maintained at the SL. Similarly, Fig. \ref{Queue_infotainment} shows the CDF of the queue length of infotainment sub-slice which maintains the video streaming queues at the RSU. From Fig. \ref{Queue_autonomous} we can see that the packet length of baseline 1 and 2 is much larger than the proposed network slicing algorithm for the autonomous driving sub-slice. The packet queue length decreases when the number of vehicles in the network decreases. The proposed network slicing algorithm with $\sigma = 5$ m and inter-vehicular distance of $200-300$ m achieves the queue length of 1 packet for 99 \% of the vehicles which indicates that all the packets are served as they arrive. Moreover, Fig. \ref{Queue_infotainment} plots the CDF of queue length of the infotainment sub-slice, where the performance of the proposed algorithm in terms of the queue length is a compromise between the baselines. 

\begin{figure}[hbtp]
	\centering
	\includegraphics[width=0.48\textwidth]{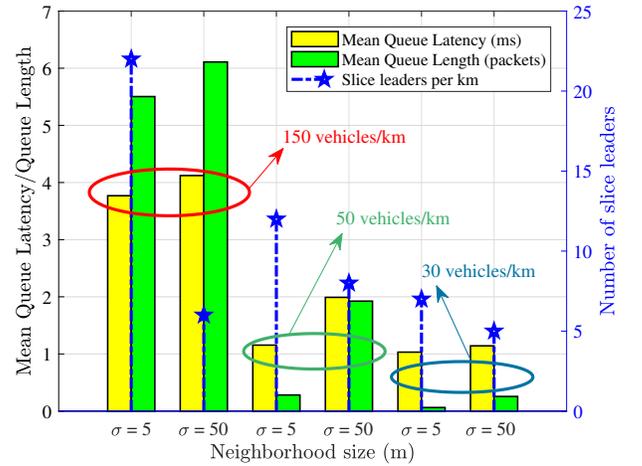}
	\caption{\label{mean} Average queue length and queuing latency for different vehicular densities.}
\end{figure}

Furthermore, we analyze the impact of neighborhood size on the number of SLs, the average queue length and the queuing latency in Fig. \ref{mean}. The neighborhood size parameter effects the similarity of vehicles i.e., small value of neighborhood size results into more number of SLs with less number of clustered vehicles per SL and vice-versa. When the network is dense the neighborhood size of 5 m results into 22 SLs per km, which is reduced to 6 SLs when neighborhood size is increased to 50m for the same scenario. When the number of SLs in the network is small and the number of clustered vehicles under each SL is large, we observe an increase in queue length and queue latency due to congestion at the SL. On the other hand, sparse network with vehicular density of 30 vehicles/km achieves the queuing latency of about 1ms, which means that the packets are served without any delay. From the Fig. \ref{mean} we can observe that smaller value of neighborhood size are preferred as they yield the minimum latency and minimum queue lengths for all the vehicular densities. 

\vspace{-5mm}

\section{Conclusion}
In this work we have studied the performance of the proposed network slicing in a vehicular environment. The scenario consist of a six lane highway layout and models different vehicular densities. The performance of the proposed method is analyzed in terms of the queuing latency, throughput and the queue lengths using an LTE-A compliant system level simulator ,with enhancement of C-V2X standard. The proposed network slicing algorithm involves the re-configuration of vehicular UE functionality to provide flexibility and scalability. We studied two sub-slices in this work to model heterogeneous traffic where the safety message traffic is mission critical and requires ultra reliable and low latency connectivity and the infotainment slice requires higher data rate. In terms of achieving the reliability the proposed network slicing algorithm outperforms both the baselines. On the other hand, the proposed technique performs better than the baseline 1 for the infotainment slice. Simulation result shows the proposed network slicing approach attains huge improvements in terms of reliability and throughput, which is due to the utilization of high quality V2V and V2I links.

\section*{Acknowledgment}
This research has been supported by the Academy of Finland 6Genesis Flagship project and the Thule institute strategic project SAFARI.
	
\bibliographystyle{IEEEtran}

\begin{thebibliography}{10}
\providecommand{\url}[1]{#1}
\csname url@samestyle\endcsname
\providecommand{\newblock}{\relax}
\providecommand{\bibinfo}[2]{#2}
\providecommand{\BIBentrySTDinterwordspacing}{\spaceskip=0pt\relax}
\providecommand{\BIBentryALTinterwordstretchfactor}{4}
\providecommand{\BIBentryALTinterwordspacing}{\spaceskip=\fontdimen2\font plus
\BIBentryALTinterwordstretchfactor\fontdimen3\font minus
  \fontdimen4\font\relax}
\providecommand{\BIBforeignlanguage}[2]{{%
\expandafter\ifx\csname l@#1\endcsname\relax
\typeout{** WARNING: IEEEtran.bst: No hyphenation pattern has been}%
\typeout{** loaded for the language `#1'. Using the pattern for}%
\typeout{** the default language instead.}%
\else
\language=\csname l@#1\endcsname
\fi
#2}}
\providecommand{\BIBdecl}{\relax}
\BIBdecl

\bibitem{ultra}
G.~Araniti, C.~Campolo, M.~Condoluci, A.~Iera, and A.~Molinaro, ``{LTE} for
  vehicular networking: a survey,'' \emph{IEEE Communications Magazine},
  vol.~51, no.~5, pp. 148--157, May 2013.

\bibitem{national}
{National Highway Traffic Safety Administration}, ``National motor vehicle
  crash causation survey,''
  \url{https://crashstats.nhtsa.dot.gov/Api/Public/ViewPublication/811059},
  report no. DOT HS 811 059, 2008.

\bibitem{V2X_IEEE}
L.~Le, A.~Festag, R.~Baldessari, and W.~Zhang, ``Vehicular wireless short-range
  communication for improving intersection safety,'' \emph{IEEE Communications
  Magazine}, vol.~47, no.~11, pp. 104--110, Nov 2009.

\bibitem{unbounded}
A.~Vinel, ``{3GPP LTE} versus {IEEE 802.11p/WAVE}: Which technology is able to
  support cooperative vehicular safety applications?'' \emph{IEEE Wireless
  Communications Letters}, vol.~1, no.~2, pp. 125--128, Apr 2012.

\bibitem{V2X_LTE2}
C.~Lottermann, M.~Botsov, P.~Fertl, and R.~Mullner, ``Performance evaluation of
  automotive off-board applications in lte deployments,'' in \emph{2012 IEEE
  Vehicular Networking Conference (VNC)}, Nov 2012, pp. 211--218.

\bibitem{speed}
{ETSI TS 122 185}, ``{LTE}; service requirements for {V2X} services ({3GPP TS
  22.185 version 14.3.0 Release 14}),''
  \url{http://www.etsi.org/deliver/etsi_en/302600_302699/30263702/01.03.00_20/en_30263702v010300a.pdf},
  Mar 2017.

\bibitem{ref7}
{3GPP TR 22.886 V1.0.0:}, ``Study on enhancement of 3{GPP} support for 5{G}
  {V2X} services ({R}elease 15),'' 2016.

\bibitem{RRM}
\BIBentryALTinterwordspacing
A.~Anpalagan, M.~Bennis, and R.~Vannithamby, \emph{Design and Deployment of
  Small Cell Networks}.\hskip 1em plus 0.5em minus 0.4em\relax Cambridge
  University Press, 2015. [Online]. Available:
  \url{https://books.google.fi/books?id=ZjoACwAAQBAJ}
\BIBentrySTDinterwordspacing

\bibitem{mehdi}
M.~{Bennis}, M.~{Debbah}, and H.~V. {Poor}, ``Ultrareliable and low-latency
  wireless communication: Tail, risk, and scale,'' \emph{Proceedings of the
  IEEE}, vol. 106, no.~10, pp. 1834--1853, Oct 2018.

\bibitem{noma}
C.~{Chen}, B.~{Wang}, and R.~{Zhang}, ``Interference hypergraph-based resource
  allocation (ihg-ra) for noma-integrated v2x networks,'' \emph{IEEE Internet
  of Things Journal}, vol.~6, no.~1, pp. 161--170, Feb 2019.

\bibitem{noma1}
B.~{Di}, L.~{Song}, Y.~{Li}, and Z.~{Han}, ``V2x meets noma: Non-orthogonal
  multiple access for 5g-enabled vehicular networks,'' \emph{IEEE Wireless
  Communications}, vol.~24, no.~6, pp. 14--21, Dec 2017.

\bibitem{resource_power}
A.~{Masmoudi}, S.~{Feki}, K.~{Mnif}, and F.~{Zarai}, ``Efficient radio resource
  management for d2d-based lte-v2x communications,'' in \emph{2018 IEEE/ACS
  15th International Conference on Computer Systems and Applications (AICCSA)},
  Oct 2018, pp. 1--6.

\bibitem{resource_1}
W.~{Sun}, D.~{Yuan}, E.~G. {Ström}, and F.~{Brännström}, ``Cluster-based
  radio resource management for d2d-supported safety-critical v2x
  communications,'' \emph{IEEE Transactions on Wireless Communications},
  vol.~15, no.~4, pp. 2756--2769, April 2016.

\bibitem{resource_2}
P.~{Wang}, B.~{Di}, H.~{Zhang}, K.~{Bian}, and L.~{Song}, ``Cellular v2x
  communications in unlicensed spectrum: Harmonious coexistence with vanet in
  5g systems,'' \emph{IEEE Transactions on Wireless Communications}, vol.~17,
  no.~8, pp. 5212--5224, Aug 2018.

\bibitem{network_slicing}
S.~A.~A. {Shah}, E.~{Ahmed}, M.~{Imran}, and S.~{Zeadally}, ``5g for vehicular
  communications,'' \emph{IEEE Communications Magazine}, vol.~56, no.~1, pp.
  111--117, Jan 2018.

\bibitem{claudia}
C.~Campolo, A.~Molinaro, A.~Iera, and F.~Menichella, ``{5G} network slicing for
  vehicle-to-everything services,'' \emph{IEEE Wireless Communications},
  vol.~24, no.~6, pp. 38--45, Dec 2017.

\bibitem{ran_slicing}
H.~D. {Albonda} and J.~{Pérez-Romero}, ``An efficient ran slicing strategy for
  a heterogeneous network with embb and v2x services,'' \emph{IEEE Access}, pp.
  1--1, 2019.

\bibitem{claudia_1}
C.~{Campolo}, A.~{Molinaro}, A.~{Iera}, R.~R. {Fontes}, and C.~E. {Rothenberg},
  ``Towards 5g network slicing for the v2x ecosystem,'' in \emph{2018 4th IEEE
  Conference on Network Softwarization and Workshops (NetSoft)}, June 2018, pp.
  400--405.

\bibitem{3gpp_NS}
{3GPP TR 23.786 v0.8.0}, ``Study on architecture enhancements for {EPS} and
  5{G} system to support advanced {V2X} services services; ({R}elease 16),''
  {F}rance, 2018.

\bibitem{slice1}
P.~Rost, C.~Mannweiler, D.~S. Michalopoulos, C.~Sartori, V.~Sciancalepore,
  N.~Sastry, O.~Holland, S.~Tayade, B.~Han, D.~Bega, D.~Aziz, and H.~Bakker,
  ``Network slicing to enable scalability and flexibility in 5{G} mobile
  networks,'' \emph{IEEE Communications Magazine}, vol.~55, no.~5, pp. 72--79,
  May 2017.

\bibitem{metis_slicing}
\BIBentryALTinterwordspacing
``{METIS} deliverable {D}7.3 {F}inal 5{G} visualization,'' Jun 2017. [Online].
  Available:
  \url{https://metis-ii.5g-ppp.eu/wp-content/uploads/deliverables/METIS-II_D7.3_V1.0.pdf}
\BIBentrySTDinterwordspacing

\bibitem{99cite}
{3GPP TS 22.185 v14.3.0}, ``{LTE}; service requirements for {V2X} services
  ({Release 14}),'' Mar 2017.

\bibitem{WINII}
\BIBentryALTinterwordspacing
``{WINNER II channel models, D1.1.2 V1.0}.'' [Online]. Available:
  \url{http://www.cept.org/files/1050/documents/winner2-finalreport.pdf}
\BIBentrySTDinterwordspacing

\bibitem{V2V_pathloss}
J.~Karedal, N.~Czink, A.~Paier, F.~Tufvesson, and A.~F. Molisch, ``Path loss
  modeling for vehicle-to-vehicle communications,'' \emph{IEEE Transactions on
  Vehicular Technology}, vol.~60, no.~1, pp. 323--328, Jan 2011.

\bibitem{sdn}
D.~{Kreutz}, F.~M.~V. {Ramos}, P.~E. {Veríssimo}, C.~E. {Rothenberg},
  S.~{Azodolmolky}, and S.~{Uhlig}, ``Software-defined networking: A
  comprehensive survey,'' \emph{Proceedings of the IEEE}, vol. 103, no.~1, pp.
  14--76, Jan 2015.

\bibitem{clustering_tutorial}
\BIBentryALTinterwordspacing
U.~von Luxburg, ``A tutorial on spectral clustering,'' \emph{CoRR}, vol.
  abs/0711.0189, 2007. [Online]. Available:
  \url{http://arxiv.org/abs/0711.0189}
\BIBentrySTDinterwordspacing

\bibitem{gaussian}
A.~L. Yuille and N.~M. Grzywacz, ``The motion coherence theory,'' in
  \emph{[1988 Proceedings] Second International Conference on Computer Vision},
  Dec 1988, pp. 344--353.

\bibitem{hamza}
H.~Khan, P.~Luoto, M.~Bennis, and M.~Latva-aho, ``On the application of network
  slicing for 5{G-V2X},'' in \emph{European Wireless 2018; 24th European
  Wireless Conference}, May 2018, pp. 1--6.

\bibitem{3GPP_frequency}
\BIBentryALTinterwordspacing
``{3GPP TR 36.785 V14.0.0}. {Vehicle to Vehicle {(V2V)} services based on LTE
  sidelink; User Equipment {(UE)} radio transmission and reception }.''
  [Online]. Available: \url{http://www.3gpp.org/dynareport/36785.htm}
\BIBentrySTDinterwordspacing

\bibitem{Cv2x}
{3GPP TR 36.300 v14.3.0}, ``Evolved universal terrestrial radio access
  {(E-UTRA)} and evolved universal terrestrial radio access network
  {(E-UTRAN)}; ({R}elease 14),'' 2017.

\bibitem{Petri2}
P.~Luoto, M.~Bennis, P.~Pirinen, S.~Samarakoon, K.~Horneman, and M.~Latva-aho,
  ``Vehicle clustering for improving enhanced {LTE-V2X} network performance,''
  in \emph{2017 European Conference on Networks and Communications (EuCNC)},
  Oulu, Finland, Jun 2017, pp. 1--5.

\end{thebibliography}

\end{document}